\UseRawInputEncoding
\documentclass[aps,pra,twocolumn]{revtex4}

\usepackage{amsmath}
\usepackage{amssymb}
\usepackage{mathrsfs}
\usepackage{epsfig}
\usepackage{color}
\usepackage[bookmarks]{hyperref}
\usepackage{graphicx}
\usepackage{bm}
\usepackage{times}
\usepackage{soul}
\usepackage{caption}
\usepackage{subcaption}

\graphicspath{{.}}

\newcommand*{\bra}[1]{\ensuremath{\langle{#1}\rvert}}
\newcommand*{\ket}[1]{\ensuremath{\lvert{#1}\rangle}}
\newcommand{\rmd}{\ensuremath{\mathrm{d}}}

\let\Re\undefined
\let\Im\undefined
\DeclareMathOperator{\Tr}{tr}
\DeclareMathOperator{\Re}{Re}
\DeclareMathOperator{\Im}{Im}

\newcommand{\eref}[1]{Eq.~\eqref{eq:#1}}
\newcommand{\erefs}[1]{Eqs.~\eqref{eq:#1}}
\newcommand{\Fref}[1]{Figure~\ref{fig:#1}}
\newcommand{\fref}[1]{Fig.~\ref{fig:#1}}
\newcommand{\frefs}[1]{Figs.~\ref{fig:#1}}
\newcommand{\sref}[1]{Sec.~\ref{sec:#1}}
\newcommand{\aref}[1]{App.~\ref{sec:#1}}

\begin{document}

\title{Optimal estimation of the optomechanical coupling strength}

\author{J\'ozsef Zsolt Bern\'ad}
\email{zsolt.bernad@um.edu.mt}
\affiliation{Department of Physics, University of Malta, Msida MSD 2080, Malta}
\author{Claudio Sanavio}
\affiliation{Department of Physics, University of Malta, Msida MSD 2080, Malta}
\author{Andr\'e Xuereb}
\affiliation{Department of Physics, University of Malta, Msida MSD 2080, Malta}

\date{\today}

\begin{abstract}
We apply the formalism of quantum estimation theory to obtain information about the value of the nonlinear optomechanical coupling strength. 
In particular, we discuss the minimum mean-square error estimator and a quantum Cram\'er--Rao-type inequality for the estimation of the coupling strength. 
Our estimation strategy reveals some cases where quantum statistical inference is inconclusive and merely result in the reinforcement of prior expectations. 
We show that these situations also involve the highest expected information losses. We demonstrate that interaction times in the order of one time period of mechanical oscillations 
are the most suitable for our estimation scenario, and compare situations involving different photon and phonon excitations.
\end{abstract}

\maketitle

\section{Introduction}
Quantum estimation theory attempts to find the best strategy for learning the value of one or more parameters of the density matrix of a quantum mechanical system \cite{Helstrom1969}, 
expressed as a positive-operator valued measure (POVM). An estimation protocol consists in identifying this POVM, applying its elements in repeated measurements of the system, and finally 
estimating the unknown parameters from the data set. The optimum strategy consists of those POVMs which minimize an average cost functional, typically considered for the maximum likelihood or 
mean-square error estimators. The mathematical framework for studying the conditions under which solutions of the optimization problem exists was established by Holevo~\cite{Holevo1973, Holevo2011}. 
Subsequently, considerable theoretical and experimental developments in quantum statistical inference have led to various applications in quantum tomography and metrology~\cite{Paris2004}.

We explore the implementation of this methodology to the specific case of an optomechanical system. It has been known for a long time that electromagnetic radiation exerts ``radiation pressure'' on 
any surface exposed to it~\cite{Aspelmeyer2014}. The resulting momentum transfer is particularly notable on mirrors forming a cavity for the electromagnetic radiation. This subject has been brought 
to the forefront in recent years because of its impact on the design and operation on laser-based interferometric gravitational wave observatories~\cite{Abbott2016}, where it imposes limits on the 
continuous detection of the mirror positions~\cite{Braginski1975,Caves1981}. Under the guise of optomechanics, it has also drawn much attention both theoretically and experimentally~\cite{Aspelmeyer2014}, 
motivated by its applications in sensitive optical detection of weak forces, mechanical motion in the quantum regime, and coherent light--matter interfaces that could form the backbone of future 
quantum information devices.

The simplest optomechanical interaction is deceptively simple to describe. Starting from two uncoupled harmonic oscillators, one representing a single mode of the electromagnetic field and one representing 
the mechanical 
oscillator, the only other free parameter in this model is the coupling strength that quantifies the extent by which the mirror rest position moves when a single photon is added to the electromagnetic field. 
The aim of this paper is to obtain the optimal strategy for measuring the value of this coupling strength. Our approach is based on quantum inference techniques, which have been successfully applied to phase 
estimations of quantum states~\cite{Giovannetti2011, Berni2015}. In our case the parameter to be estimated is not a simple phase parameter, but rather a parameter which appears both in the 
spectrum~\cite{Hayashi2006,Seveso2017} and the eigenvalues of the quantum state. If the motion of the mechanical oscillator is adiabatically slow compared to the frequency separation of the optical 
modes~\cite{Law1995}, an analytical solution can be found for the evolution of a system described by this model~\cite{Bernad2013,Bose1997}. The resulting density matrix describes the joint state of the 
field and the mechanical oscillator. Since measurements are, in practice, usually performed on the state of the electromagnetic field emerging from the system, we trace out the mechanical degrees of freedom. 
We therefore concentrate on the resulting optical state, which will be subject of our quantum estimation procedure.

In this paper we shall focus on a mean-square error estimator and assume that the prior probability density function of the coupling constant is a Gaussian distribution, which will allow us to keep our 
calculations analytic as far as possible. We set the mean and the standard deviation of this distribution function to values obtained by a canonical quantization procedure with a high frequency cut-off of 
the radiation field and adiabatically slow motion of the mechanical oscillator. In order to illustrate the basic features of our proposal, we consider the mechanical oscillator to be initially in 
(i)~a coherent state, (ii)~a thermal state, and (iii)~a squeezed state. For the sake of simplicity and to keep our numerical calculations tractable, we will assume that the initial optical field state has 
only a few excitations. We determine the mean-square error estimator which minimizes the cost functional and study a quantum Cram\'er--Rao-type inequality for the mean-squared error of a biased estimator. 
For our model, a right logarithmic or a symmetrized logarithmic derivative operator of the electromagnetic field state with respect to the coupling constant is very challenging to construct since the Hilbert space 
of the harmonic oscillator is infinite-dimensional. Therefore, we explore the possibilities of deriving an analytically accessible lower bound formula of the mean-squared error of the estimator. Obviously, 
we will employ the standard techniques for the derivation of the unbiased quantum Cram\'er--Rao inequality~\cite{Helstrom1974}.

This paper is organized as follows. In \sref{model} we discuss the model and its solutions of the single mode radiation coupled via radiation pressure to a vibrational mode of a mechanical oscillator. 
In \sref{mst} we introduce quantum estimation theory for minimizing mean-square error estimators, and study its properties when applied to the optomechanical model. We then address the mean-squared error of the 
biased estimator, in \sref{qCR}, deriving a lower bound and employing this result to the optomechanical model. A discussion about our analytical and numerical findings is the subject of a summary in 
\sref{conclusion}.

\section{Model}
\label{sec:model}
We consider a system composed of two harmonic oscillators, a single mode of the radiation field and a vibrational mode of a mechanical oscillator. Provided that the electromagnetic field 
describes a high-finesse cavity field mode, and that one of the mirrors is movable, one is able to derive a radiation-pressure interaction Hamiltonian~\cite{Law1994, Law1995} by using time-varying 
boundary conditions in the quantization procedure, resulting in ($\hbar=1$)
\begin{equation}
\hat{H}=\omega_\text{c} \hat{a}^\dagger \hat{a} + \omega_\text{m} \hat{b}^\dagger \hat{b}+ g \hat{a}^\dagger \hat{a} (\hat{b}^\dagger+ \hat{b}),
\label{eq:optoH}
\end{equation}
where $\hat{a}$ ($\hat{a}^\dagger$) is the annihilation (creation) operator of the single mode electromagnetic field, with frequency $\omega_\text{c}$, and $\hat{b}$ ($\hat{b}^\dagger$) is the 
annihilation (creation) operator of the mirror motion, with frequency $\omega_\text{m}$. The strength of the optomechanical interaction, denoted $g$, depends on the specifics of the realization in question.

Starting from a joint field--mechanics state $\ket{\Psi(0)}$, the time evolution of the system is given by the Schr\"odinger equation and can be rephrased as
\begin{equation}
 \ket{\Psi(t)}=e^{-i \hat{H} t} \ket{\Psi(0)}.
\end{equation}
We are interested in the case where there are no initial correlations between the field and the mechanical oscillator. Therefore, we choose an initial state of the form
\begin{equation}
\ket{\Psi(0)}=\sum^\infty_{n=0} a_n \ket{n}_\text{c} \ket{\psi}_\text{m},
\label{eq:initial}
\end{equation}
where the exact form of $\ket{\psi}_\text{m}$ depends on the desired initial conditions. In the following subsections we consider initial coherent, thermal, and squeezed states.

\subsection{Coherent state}
We set
We start off by setting the initial mechanical oscillator to a coherent state \cite{Perelomov1986}
\begin{equation}
 \ket{\psi}_\text{m}=\ket{\alpha}_\text{m}=\sum_{n=0}^\infty
 e^{-\frac{\lvert\alpha\rvert^2}{2}}
 \frac{\alpha^n}{\sqrt{n!}}
 \ket{n}_\text{m},\ \text{with}\ \alpha=\lvert\alpha\rvert\,e^{i\phi},
\label{eq:cohinitial}
\end{equation}
which we write in terms of the field number states $\ket{n}_\text{m}$ ($n\in {\mathbb N}_0$). Here, $\lvert\alpha\rvert$ is the amplitude of the coherent state and $\phi$ its phase. 
We allow the coefficients $a_n$ of the photon-number states to be general and only impose the normalization condition $\sum_n\lvert{a_n}\rvert^2=1$. The choice of the initial state in \eref{cohinitial} 
is our basic approach for determining the time evolution of the system, and will eventually be extended to cover initial thermal and squeezed states of the mechanical oscillator.

The interaction Hamiltonian $g \hat{a}^\dagger \hat{a} (\hat{b}^\dagger+ \hat{b})$ commutes with the free Hamiltonian of the radiation field, $\omega_\text{c}\hat{a}^\dagger \hat{a}$, which yields
\begin{equation}
 ~_\text{c}\bra{n} \hat{H} \ket{m}_\text{c}=\bigl[n \omega_\text{c} \hat{I} + \omega_\text{m} \hat{b}^\dagger \hat{b}+ n g (\hat{b}^\dagger+ \hat{b})\bigr] \delta_{n,m}
\label{eq:calc1}
\end{equation}
with $\delta_{n,m}$ being the Kronecker delta and $\hat{I}$ the identity operator on the Hilbert space of the mechanical oscillator. Thus, the Hamiltonian~\eqref{eq:optoH} is block-diagonal with 
respect to photon-number states $\ket{n}_\text{c}$.

In order to evaluate the expression $\exp \{-i \omega_\text{m} \hat{b}^\dagger \hat{b}t-i n g (\hat{b}^\dagger+ \hat{b})t\}$ we employ the Baker--Campbell--Hausdorff formula and obtain 
(see, for example, Ref.~\cite{Bernad2013})
\begin{equation}
 e^{-i \omega_\text{m} \hat{b}^\dagger \hat{b} t - i n g (\hat{b}^\dagger+ \hat{b})t} e^{i \omega_\text{m} \hat{b}^\dagger \hat{b} t}=
 e^{i\Phi_n(t)} e^{\alpha_n(t)\hat{b}^\dagger-\alpha^\ast_n(t)\hat{b}},
\label{eq:calc2}
\end{equation}
where we have introduced the parameters
\begin{align}
 \alpha_n(t)&=\frac{n g}{\omega_\text{m}}\bigl(e^{-i\omega_\text{m} t}-1\bigr)\ \text{and}\\
 \Phi_n(t)&=\frac{n^2 g^2}{\omega^2_\text{m}}\bigl[\omega_\text{m} t-\sin(\omega_\text{m} t)\bigr].
\end{align}
This implies that the full time evolution can be viewed as photon-number dependent displacements of the mechanical oscillator; with the help of \erefs{calc1} and~\eqref{eq:calc2}, we find
\begin{subequations}
\begin{align}
\ket{\Psi(t)}&=\sum^\infty_{n=0} a_n e^{i \varphi_n(t)} \ket{n}_\text{c}
 \ket{\beta_n(t)}_\text{m},\\
 \varphi_n(t)&=-n \omega_\text{c} t + \frac{n^2 g^2}{\omega^2_\text{m}}\bigl[\omega_\text{m} t-\sin(\omega_\text{m} t)\bigr]\\
 &\qquad+\frac{ng}{\omega_\text{m}}
 \frac{\alpha^\ast{}(1-e^{i\omega_\text{m} t})-\alpha(1-e^{-i\omega_\text{m} t}) }{2i},\ \text{and}\nonumber\\
 \beta_n(t)&=\frac{n g}{\omega_\text{m}}\bigl(e^{-i\omega_\text{m} t}-1\bigr)+ \alpha e^{-i \omega_\text{m} t},
\end{align}
\label{eq:mainr}
\end{subequations}
where we have also used a corollary of the Baker--Campbell--Hausdorff formula, which states that the product of two displacement operators is also a displacement operator with an overall phase factor.

The quantum state of \eref{mainr} yields a complete description of the interaction between the single mode of the radiation field and the single vibration mode of the mechanical oscillator, i.e., 
neglecting all losses and sources of decoherence. In the subsequent sections we will be interested in possible measurement scenarios of the field, which are capable to estimate the couple constant $g$. 
Therefore, the field to be measured for estimation reads
\begin{equation}
 \hat{\rho}_\text{F}=\Tr_{\text{mech}} \{\ket{\Psi(t)}\bra{\Psi(t)}\}=\sum^\infty_{n,m=0} A_{n,m} \ket{n}_\text{c} \bra{m}
\label{eq:densitym}
\end{equation}
with
\begin{align}
 A_{n,m}&=a_n a^\ast_\text{m} e^{i \varphi_n(t)-i\varphi_\text{m}(t)-\left(|\beta_n(t)|^2+|\beta_\text{m}(t)|^2\right)/2+\beta_n(t)\beta^\ast_\text{m}(t)}\nonumber\\
 &=a_n a^\ast_\text{m} e^{-g^2 f^{(2)}_{n,m}(t)+g f^{(1)}_{n,m}(t)-f^{(0)}_{n,m}(t)},
\label{eq:Aform}
\end{align}
where
\begin{subequations}
\begin{align}
f^{(0)}_{n,m}(t)&=i \omega_\text{c} t (n-m),\\
f^{(1)}_{n,m}(t)&=\frac{\alpha^\ast{}(1-e^{i\omega_\text{m} t})-\alpha(1-e^{-i\omega_\text{m} t}) }{\omega_\text{m}}(n-m),\ \text{and}\\
f^{(2)}_{n,m}(t)&=\frac{1-\cos(\omega_\text{m} t)}{\omega^2_\text{m}} (n-m)^2\nonumber\\
&\qquad-i \frac{\omega_\text{m} t-\sin(\omega_\text{m} t)}{\omega^2_\text{m}} (n^2-m^2).
\end{align}
\label{eq:example1}
\end{subequations}
We note in passing that these results imply that the coefficient of the linear term in $g$ contributes to $A_{n,m}$ only when the initial state of the mechanical oscillator is not in the ground state, 
i.e., $\alpha\neq0$. Before moving on to developing the quantum error theory we require, let us briefly extend our considerations to thermal and squeezed states.

\subsection{Thermal state}

If the initial state of the mechanical oscillator is thermal, as a first step we must switch from discussing state vectors to density matrices. In this case, the uncorrelated initial state of the 
optomechanical system has the form
\begin{equation}
\hat{\rho}(t=0)=\sum^\infty_{n,m=0} a_n a^\ast_\text{m} \int \frac{e^{-\lvert\gamma\rvert^2/n_{\text{th}}}}{\pi n_{\text{th}}} \ket{n}_\text{c} \ket{\gamma}_\text{m} \bra{m}_\text{c} \bra{\gamma}_\text{m}\,
\rmd^2 \gamma,
\end{equation}
where we have used the Glauber--Sudarshan representation \cite{Glauber1963, Sudarshan1963} of the mechanical oscillator thermal state with the average phonon number
\begin{equation}
 n_{\text{th}} = \biggl[\exp \biggl( \frac{\hbar\omega_\text{m}}{k_\text{B}T} \biggr)-1 \biggr]^{-1},
\end{equation}
where $k_\text{B}$ is the Boltzmann constant and $T$ the thermodynamic temperature of the initial state of the mechanical system. The time evolution of this system is given by
\begin{equation}
 \hat{\rho}(t)=e^{-i \hat{H} t} \hat{\rho}(0) e^{i \hat{H} t},
\end{equation}
which yields
\begin{multline}
 \hat{\rho}(t)=\sum^\infty_{n,m=0} a_n a^\ast_\text{m} \int \frac{e^{-\lvert\gamma\rvert^2/n_{\text{th}}}}{\pi n_{\text{th}}} e^{i \varphi_n(t)-i \varphi_\text{m}(t)}
 \ket{n}_\text{c} \bra{m}_\text{c}\\
 \times\ket{\beta_n(t)}_\text{m} \bra{\beta_\text{m}(t)}_\text{m} \,\rmd^2 \gamma,
\end{multline}
with
\begin{align}
 \varphi_n(t)&=-n \omega_\text{c} t + \frac{n^2 g^2}{\omega^2_\text{m}}\left[\omega_\text{m} t-\sin(\omega_\text{m} t)\right]\nonumber\\
 &\qquad+\frac{ng}{\omega_\text{m}}
 \frac{\gamma^\ast{}(1-e^{i\omega_\text{m} t})-\gamma(1-e^{-i\omega_\text{m} t}) }{2i},\ \text{and}\\
 \beta_n(t)&=\frac{n g}{\omega_\text{m}}\left(e^{-i\omega_\text{m} t}-1\right)+ \gamma e^{-i \omega_\text{m} t}.
\end{align}
In the next step we trace out the mechanical degrees of freedom, as before, obtaining
\begin{multline}
 \hat{\rho}_\text{F}=\Tr_{\text{mech}} \{\hat{\rho}(t)\}=\sum^\infty_{n,m=0} a_n a^\ast_\text{m} \ket{n}_\text{c} \bra{m}_\text{c}\\
 \times \int \frac{e^{-\lvert\gamma\rvert^2/n_{\text{th}}}}{\pi n_{\text{th}}} B_{n,m}(\gamma,\gamma^\ast{}) \,\rmd^2\gamma,
\end{multline}
with
\begin{equation}
 B_{n,m}(\gamma,\gamma^\ast{})= e^{-h_0+h_1},
\end{equation}
where
\begin{align}
h_0&=i \omega_\text{c} t (n-m)+\frac{g^2}{\omega^2_\text{m}} \Big\{ \big[1-\cos(\omega_\text{m} t)\big](n-m)^2\nonumber\\
&\qquad-i \big[ \omega_\text{m} t-\sin(\omega_\text{m} t) \big] (n^2-m^2) \Big\},\ \text{and}\\
h_1&=\frac{g}{\omega_\text{m}} \big[\gamma^\ast{}(1-e^{i\omega_\text{m} t})-\gamma(1-e^{-i\omega_\text{m} t}) \big] (n-m).
\end{align}
Now, we perform the Gaussian integral by using $\rmd^2\gamma=\rmd\Re\{\gamma\}\,\rmd\Im\{\gamma\}$ and obtain a density matrix in the form of \eref{densitym}. Employing the notation of \eref{Aform}, we find
\begin{align}
f^{(0)}_{n,m}(t)&=i \omega_\text{c} t (n-m),\\
f^{(1)}_{n,m}(t)&=0,\ \text{and}\\
f^{(2)}_{n,m}(t)&=(2 n_{\text{th}}+1)\frac{1-\cos(\omega_\text{m} t)}{\omega^2_\text{m}} (n-m)^2\nonumber\\
&\qquad-i \frac{\omega_\text{m} t-\sin(\omega_\text{m} t)}{\omega^2_\text{m}} (n^2-m^2).
\label{eq:example2}
\end{align}

\subsection{Squeezed state}

Let us now consider the case where the initial state of the mechanical system is a displaced squeezed state; we write, therefore,
\begin{equation}
\ket{\Psi(0)}=\sum^\infty_{n=0} a_n \ket{n}_\text{c} \ket{\alpha,\zeta}_\text{m}
\end{equation}
with the mechanical oscillator state being defined as~\cite{Schleich2001}
\begin{equation}
\ket{\alpha,\zeta}_\text{m}=\hat{D}(\alpha)\hat{S}(\zeta)\ket{0}_\text{m}
\label{eq:squeezed}
\end{equation}
where $\hat{D}(\alpha)=\exp\bigl(\alpha\hat{b}^\dagger-\alpha^\ast\hat{b}\bigr)$, with $\alpha=\lvert\alpha\rvert{}e^{i\phi}$, is the displacement operator, and $\hat{S}(\zeta)=\exp\bigl[\tfrac{1}{2}
\bigl(\zeta^\ast{}b^2-\zeta{}b^{\dagger 2}\bigr)\bigr]$, with $\zeta=\lvert\zeta\rvert{}e^{i\theta}$, is the squeezing operator.

We employ the squeezed state of \eref{squeezed}; in passing, however, we note that it is possible to invert the order of the displacement and squeezing operator. This
results in a generalized squeezed state, which differs from the original state by the displacement parameter:
\begin{equation}
\hat{S}(\zeta)\hat{D}(\alpha)=\hat{D}\bigl[\alpha\cosh(\lvert\zeta\rvert)-\alpha^\ast{} e^{i\theta} \sinh(\lvert\zeta\rvert)\bigl]\hat{S}(\zeta).
\end{equation}
Exploiting the block-diagonal structure of the Hamiltonian with respect to the photon-number states $\ket{n}_\text{c}$, and \eref{calc2}, we find
\begin{equation}
~_\text{c}\bra{n} e^{-i\hat{H}t} \ket{\Psi(0)}=a_n e^{i \varphi_n(t)} \ket{\beta_n(t),\zeta{}e^{-2i\omega_\text{m} t}}_\text{m}
\end{equation}
where $\varphi_n(t)$ and $\beta_n(t)$ are defined in \erefs{mainr}. Next, tracing out the mechanical degrees of freedom yields the state of the field in the form of \eref{densitym}, with
\begin{multline}
 A_{n,m}=a_n a^\ast_m e^{i \varphi_n(t)-i\varphi_\text{m}(t)}\\
 \times\Tr\{\ket{\beta_n(t),\zeta{}e^{-2i\omega_\text{m} t}}\bra{\beta_m(t),\zeta{}e^{-2i\omega_\text{m} t}}\}.
\end{multline}
\begin{widetext}
The trace in this equation can be evaluated with the help of the Glauber--Sudarshan representation, which allows us to write
\begin{equation}
\Tr\{\ket{\beta_n(t),\zeta{}e^{-2i\omega_\text{m} t}}\bra{\beta_m(t),\zeta{}e^{-2i\omega_\text{m} t}}\}=\int \frac{\rmd^2\gamma}{\pi}\langle\gamma \ket{\beta_n(t),\zeta{}e^{-2i\omega_\text{m} t}}
\bra{\beta_m(t),\zeta{}e^{-2i\omega_\text{m} t}} \gamma \rangle.
\label{eq:tracetodosq}
\end{equation}
First, we note that
\begin{equation}
\langle\gamma \ket{\beta_n(t),\zeta{}e^{-2i\omega_\text{m} t}}_\text{m}=e^{-(\gamma \beta_n^\ast{}(t)-\gamma^\ast{}\beta_n(t))/2}\bra{0}\hat{D}^\dagger\bigl(\gamma-\beta_n(t)\bigr)\hat{S}(\zeta{}
e^{-2i\omega_\text{m} t})\ket{0},
\label{eq:todosq}
\end{equation}
where we have used the relation $\hat{D}^\dagger(-\gamma)=\hat{D}(\gamma)$. The overlap integral between the coherent state $\ket{\gamma}$ and the squeezed state $\ket{0,\zeta}$ is
\begin{equation}
\langle \gamma \ket{0,\zeta}=\sqrt{\frac{e^{-\lvert\gamma\rvert^2}}{\cosh(\lvert\zeta\rvert)}} \exp\bigl[-\tfrac{1}{2} \gamma^{\ast2}e^{i\theta}\tanh(\lvert\zeta\rvert)\bigr].
\end{equation}
For the purposes of \eref{todosq} we thus obtain
\begin{equation}
\langle\gamma \ket{\beta_n(t),\zeta{}e^{-2i\omega_\text{m} t}}_\text{m}=\sqrt{\frac{e^{-|\gamma-\beta_n(t)|^2-\gamma \beta_n^\ast{}(t)+\gamma^\ast{}\beta_n(t)}}{\cosh(\lvert\zeta\rvert)}}
\exp\bigl\{-\tfrac{1}{2}\bigl[\gamma^\ast{}-\beta^\ast_n(t)\bigr]^{2}e^{i(\theta-2 \omega_\text{m} t)}\tanh(\lvert\zeta\rvert)\bigr\}.
\end{equation}
Substituting this result into \eref{tracetodosq} and performing the integral by using $\rmd^2\gamma=\rmd\Re\{\gamma\}\,\rmd\Im\{\gamma\}$
we obtain the coefficients in \eref{Aform}:
\begin{subequations}
\begin{align}
f^{(0)}_{n,m}&=i \omega_\text{c} t(n-m)+\lvert\alpha\rvert^2\big[ 1+\tanh(\lvert\zeta\rvert)\cos{(\theta-2\phi)}\big]-I^{(0)}+\ln\Bigl[\cosh(\lvert\zeta\rvert)\sqrt{1-\tanh^2(\lvert\zeta\rvert)}\Bigr],\\
f^{(1)}_{n,m}(t)&=\frac{\alpha^\ast{}(1-e^{i\omega_\text{m} t})}{\omega_\text{m}}(n-m)+\frac{I^{(1)}_{n,m}}{\omega_\text{m}}+\tanh{(\lvert\zeta\rvert)}\frac{\alpha^\ast{}(1-e^{-i\omega_\text{m} t})
e^{i\theta}n+\alpha(1-e^{i\omega_\text{m} t})e^{-i\theta}m}{2\omega_\text{m}}\\
f^{(2)}_{n,m}(t)&=-i \frac{\omega_\text{m} t-\sin(\omega_\text{m} t)}{\omega^2_\text{m}} (n^2-m^2)+\tanh{(\lvert\zeta\rvert)}\frac{(e^{-i\omega_\text{m} t}-1)^2e^{i\theta}n^2+(e^{i\omega_\text{m} t}-1)^2
e^{-i\theta}m^2}{2\omega^2_\text{m}}\\
&\qquad+\frac{1-\cos{\omega_\text{m} t}}{\omega^2_\text{m}} (n^2+m^2)+\frac{I^{(2)}_{n,m}}{\omega^2_\text{m}},
\end{align}
\label{eq:example3}
\end{subequations}
where, for simplicity of presentation, we relegate the explicit form of the coefficients $I^{(0)}$, $I^{(1)}_{n,m}$, and $I^{(2)}_{n,m}$ to \aref{Calc}.
\end{widetext}

\section{Quantum minimum mean-square error estimation}
\label{sec:mst}
Quantum estimation theory attempts to find the best strategy for estimating one or more parameters of the density matrix~\cite{Helstrom1976}. In our case, the density matrix of the field in \eref{densitym} 
depends on the parameter $g$ to be estimated. Any outcome of a measurement on the field is a variable with probability depending on the estimanda $g$, the parameter to be estimated. 
As our knowledge of $g$ is limited, we assume that the 
estimanda is a random variable with prior probability density function
\begin{equation}
 p(g)=\frac{1}{\sqrt{2\pi \sigma^2}} e^{-\frac{(g-g_0)^2}{2\sigma^2}}
\label{eq:apriori}
\end{equation}
with mean $g_0$ and variance $\sigma^2$. We shall return to these parameters and their physical meanings later on.

Our estimation problem is to thus find the best measurements on $\hat{\rho}_\text{F}(g)$ to estimate $g$. In practice, we are looking for a POVM the elements $\hat{\Pi}(\Delta)$ of which are defined on a compact 
interval $\Delta \subset \mathbb{R}$ of the real line, i.e., the set of all possible values for $g$, and satisfy
\begin{subequations}
\begin{equation}
 0 \leqslant \hat{\Pi}(\Delta) \leqslant \hat{I},
\end{equation}
where $\hat{I}$ is the identity operator. We also suppose that the infinitesimal operators $\rmd\hat{\Pi}(g)$ can be formed, thus yielding
\begin{equation}
 \hat{\Pi}(\Delta)=\int_{\Delta}\rmd\hat{\Pi}(g),
\end{equation}
and
\begin{equation}
 \hat{I}=\int^{\infty}_{-\infty}\rmd\hat{\Pi}(g).
\end{equation}
\label{eq:cons}
\end{subequations}
In order to solve the estimation problem we have to also provide a cost function, a measure of the cost incurred upon making errors in the estimate of $g$. Here, we wish to minimize the average squared cost of 
error, which is encoded in the cost function
\begin{equation}
 C(\tilde{g}, g)=(\tilde{g} -g)^2,
\label{eq:costf}
\end{equation}
where $\tilde{g}$ is the estimate of $g$, and therefore a function of the measurement data.

Now, we are able to formulate the quantum estimation problem. We are looking for $\rmd\hat{\Pi}(\tilde{g})$, which minimizes the average cost of this estimation strategy
\begin{equation}
 \bar C=\Tr\Bigl\{\int^{\infty}_{-\infty} \int^{\infty}_{-\infty} p(g) C(\tilde{g}, g) \hat{\rho}_\text{F}(g)\rmd\hat{\Pi}(\tilde{g}) \rmd{}g \Bigr\},
\label{eq:averagecost}
\end{equation}
under the constraints embodied in \erefs{cons}. This is a variational problem for the functional $\bar C$. To proceed, we consider each possible estimate $\tilde{g}$ to be an eigenvalue of the Hermitian operator
\begin{equation}
 \hat{M}=\int^{\infty}_{-\infty} \tilde{g}\,\rmd \hat{\Pi}(\tilde{g}) = \int^{\infty}_{-\infty} \tilde{g} \ket{\tilde{g}} \bra{\tilde{g}}\,\rmd\tilde{g}
\end{equation}
with eigenstates $\ket{\tilde{g}}$. Thus, the average cost functional in \eref{averagecost}, calculated using the cost function in \eref{costf}, can be written as
\begin{equation}
 \bar C [\hat{M}]=\Tr\Bigl\{ \int^{\infty}_{-\infty} p(g) \bigl(\hat{M}-g \hat{I}\bigr)^2 \hat{\rho}_\text{F}(g) \rmd{}g \Bigr\}.
\end{equation}
For convenience, we now define the following operators ($k=0,1,2$):
\begin{equation}
\hat{\Gamma}_k= \int^{\infty}_{-\infty} g^k p(g) \hat{\rho}_\text{F}(g)\,\rmd{}g.
\label{eq:Gamma}
\end{equation}
Now, let $\epsilon$ be a real number and $\hat{N}$ any Hermitian operator. Let $\hat{M}_{\text{min}}$ be the Hermitian operator which minimizes $\bar C [\hat{M}]$. Then, we have
\begin{equation}
 \bar C [\hat{M}_{\text{min}}] \leqslant \bar C [\hat{M}_{\text{min}}+\epsilon \hat{N}],
\end{equation}
because the sum of Hermitian operators is itself a Hermitian operator. Evaluating the right-hand side of the inequality and using the operators defined in \eref{Gamma}, we obtain
\begin{multline}
\bar C [\hat{M}_{\text{min}}+\epsilon \hat{N}]=\bar C [\hat{M}_{\text{min}}]\\
+\epsilon\,\Tr\bigl\{\hat{N}\bigl(\hat{\Gamma}_0 \hat{M}_{\text{min}} +
\hat{M}_{\text{min}} \hat{\Gamma}_0 -2\hat{\Gamma}_1 \bigr)\bigr\}\\
+\epsilon^2\,\Tr\{\hat{\Gamma}_0 \hat{N}^2 \}.
\label{eq:variat}
\end{multline}
By differentiating this relation with respect to $\epsilon$ and equating the result to zero, one is able to show that the unique Hermitian operator $\hat{M}_{\text{min}}$ minimizing $\bar C$ must 
satisfy~\cite{Personick1971}
\begin{equation}
\hat{\Gamma}_0 \hat{M}_{\text{min}} +\hat{M}_{\text{min}} \hat{\Gamma}_0=2\hat{\Gamma}_1.
\label{eq:tosolve}
\end{equation}
The average minimum cost of error for this measurement is
\begin{align}
 \bar C_{\text{min}}&=\Tr\{\hat{\Gamma}_0 \hat{M}^2_{\text{min}}-2 \hat{\Gamma}_1 \hat{M}_{\text{min}} + \hat{\Gamma}_2\}\nonumber\\
 &=\Tr\{\hat{\Gamma}_2-\hat{M}_{\text{min}} \hat{\Gamma}_0 \hat{M}_{\text{min}} \}
\label{eq:mincost},
\end{align}
where we have used the relation in \eref{tosolve} to simplify the result. In order to determine $\hat{M}_{\text{min}}$ we thus have to solve the operator equation \eref{tosolve}. It has been shown in
Ref.~\cite{Personick1971} that the unique solution of this equation can be written as
\begin{equation}
\hat{M}_{\text{min}}=2 \int^\infty_0 \exp(-\hat{\Gamma}_0 x) \hat{\Gamma}_1 \exp(-\hat{\Gamma}_0 x) \,\rmd{}x.
\label{eq:intx}
\end{equation}
A comment about this solution is in order. The operator that we have found does not necessarily represent the best estimator of $g$, but rather the measurement operator which protects best 
against information loss, no matter what the true value of $g$ is. Further discussion about this subtlety and its relation to biased estimators can be found in the illuminating monograph by Jaynes~\cite{Jaynes2003}.

We can evaluate all the $\hat{\Gamma}_k$ by using the form of $\hat{\rho}_\text{F}(g)$ in \eref{densitym}, thus obtaining ($k=0,1,2$)
\begin{equation}
\hat{\Gamma}_k=\sum^\infty_{n,m=0} a_n a^\ast_m A^{(k)}_{n,m} \exp \left(-\gamma_{n,m}\right) \ket{n}_\text{c}\bra{m},
\label{eq:gammak}
\end{equation}
with
\begin{subequations}
\begin{align}
A^{(0)}_{n,m} &= \frac{1}{\sigma'},\\
A^{(1)}_{n,m} &= \frac{g_0+f^{(1)}_{n,m}(t) \sigma^2}{\sigma'^3},\ \text{and}\\
A^{(2)}_{n,m} &= \frac{\left(g_0+f^{(1)}_{n,m}(t) \sigma^2\right)^2+\sigma^2 \sigma'^2}{\sigma'^5},
\end{align}
\end{subequations}
where we have also introduced
\begin{align}
\gamma_{n,m}&=\bigl\{2 g^2_0 f^{(2)}_{n,m}(t)- 2g_0 f^{(1)}_{n,m}(t)+2f^{(0)}_{n,m}(t)\sigma'^2\nonumber\\
&\qquad- \bigl[f^{(1)}_{n,m}(t)\bigr]^2\sigma^2\bigr\}\big/\bigl(2\sigma'^2\bigr),\ \text{and}\\
\sigma'^2&=2f^{(2)}_{n,m}(t) \sigma^2 +1.
\end{align}

Written in this form, our results are very general. In the following section we will investigate some simple cases, related to the optomechanical model introduced previously.

\section{A case study: Optomechanics}
\label{sec:OM}
We shall now put the results obtained in the two preceding sections together, thus allowing us to study the process of quantum mean-square error estimation as it applies to an optomechanical system.

The simplest non-trivial case results when $a_n=0$ for $n>1$ in \eref{initial}. In order to maximize the absolute values of the off-diagonal elements of the density matrix we choose $a_0=a_1=1/\sqrt{2}$. 
This specific choice is due to the fact that the unknown parameter $g$ is only present in the off-diagonal elements, as can be seen from \erefs{example1}. Here, the estimate $\tilde{g}$ is simply one of the 
two eigenvalues of $\hat{M}_{\text{min}}$, which turn up as a result of applying the two projective measurements defined by their accompanied eigenvectors.

Furthermore, we ought to define $g_0$ and $\sigma$ in \eref{apriori}, the prior probability density function of the estimanda $g$. We set
\begin{subequations}
\begin{align}
 g_0&=\frac{\omega_\text{c}}{L} \sqrt{\langle \hat{x}^2 \rangle_0}=\frac{\omega_\text{c}}{L} \sqrt{\frac{\hbar}{2 m \omega_\text{m}}}, \text{and}\\
 \sigma^2&=\Bigl(\frac{\omega_\text{c}}{L}\Bigr)^2 \sqrt{\langle \hat{x}^4 \rangle_0-\langle \hat{x}^2 \rangle^2_0}=\Bigl(\frac{\omega_\text{c}}{L}\Bigr)^2
 \frac{\hbar}{\sqrt{2} m \omega_\text{m}},
\end{align}
\label{eq:mst}
\end{subequations}
where $L$ is the length of the cavity, $m$ is the mass of the mechanical oscillator, and $\langle \hat{A} \rangle_0$ is the expectation value of operator $\hat{A}$, acting only on the Hilbert space of 
the mechanical oscillator, in the ground state~\cite{Aspelmeyer2014, Law1995}. For the sake of simplicity we perform our calculations in the rotating frame of the single-mode field, i.e., $\hat{\rho}_\text{F} 
\rightarrow \hat{U} \hat{\rho}_\text{F} \hat{U}^\dagger$ with $\hat{U}=\exp\{-i\omega_\text{c} t \, \hat{a}^\dagger \hat{a}\}$.

\subsection{Coherent state}

We determine $\hat{M}_{\text{min}}$ from the $\hat{\Gamma}_k$ in \eref{gammak} by using~\eref{intx}. One can obtain analytical results; however, due to their complex structure we omit their explicit 
presentation here. Instead, we focus on numerical solutions. First, we investigate the average minimum cost of error $\bar C_{\text{min}}$; \fref{costfig} shows $\bar C_{\text{min}}$ as a function of time, 
which decreases until it reaches its minimum and then returns asymptotically to its initial value, which is equal to $\sigma^2$. At $t=0$, where no interaction occurred, the eigenvalues of $\hat{M}_\text{min}$ 
are $g_0$ and zero. The probability of
measuring the eigenvalue zero is zero and therefore the estimate is $g_0$. It is immediate from the form of the prior probability distribution $p(g)$ in \eref{apriori} that the average minimum cost of error 
is $\sigma^2$, or simply the variance of $p(g)$, at $t=0$.

\begin{figure}
\includegraphics[width=0.45\textwidth]{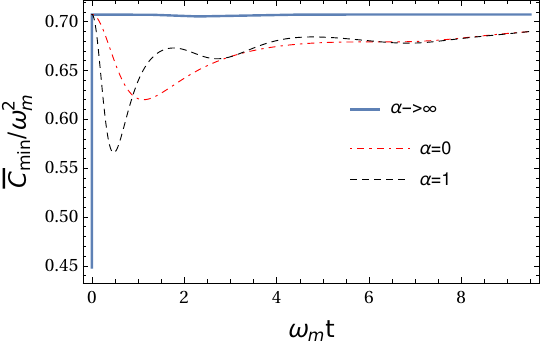}
\caption{The average minimum cost of error $\bar C_{\text{min}}/\omega^2_\text{m}$, as a function of $\omega_\text{m} t$. We consider the amplitude $\alpha$ of the initial coherent state of the mechanical 
oscillator to be real; see \erefs{example1}. We set $g_0/\omega_\text{m}=1$ and $\sigma/\omega_\text{m}=2^{-1/4}$. All curves are characterized by one global minimum which decreases with increasing $\alpha$.}
\label{fig:costfig}
\end{figure}

In the opposite limit, $t \to \infty$, the average minimum cost of error $\bar C_{\text{min}}$ is also $\sigma^2$; however, the estimates or the eigenvalues of $\hat{M}_\text{min}$ are $g_0$, 
as can be seen in \fref{eigenvM}. This means that for long interaction times the inference of the parameter $g$ from the measurement data only yields the mean $g_0$ of the probability distribution $p(g)$. 
Since the average minimum cost of error attains its maximum at both $t=0$ and $t\to\infty$, we are going to neglect these situations and focus on intermediate times, when $\bar C_{\text{min}}$ decreases. 
The fact that average minimum cost of error reaches a minimum at a finite time implies the existence of a particular duration for the interaction that yields the greatest amount of information on $g$. 
For each set of parameters, we can determine the time $t^\ast{}$ as the time when $\bar C_{\text{min}}$ reaches its minimum value. We can work backwards to obtain the specific 
$\hat{M}^\ast_\text{min}=\hat{M}_\text{min}(t=t^\ast{})$ to be measured, which is the measurement that best protects against information loss.

\begin{figure}
\includegraphics[width=0.45\textwidth]{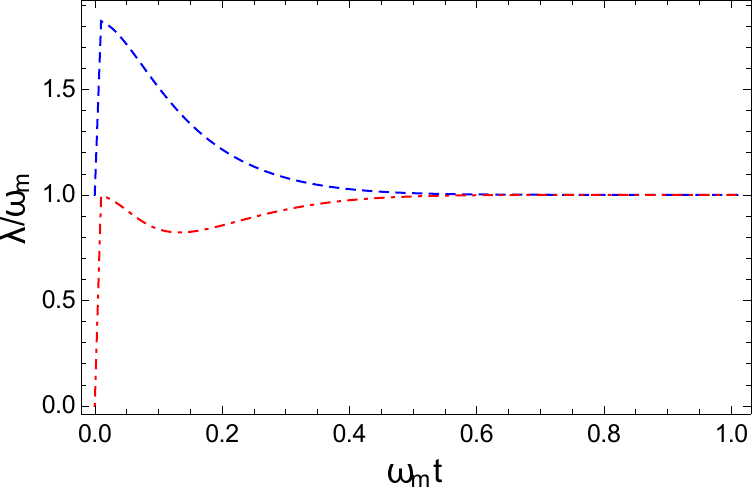}
\caption{The two eigenvalues of the operator $\hat{M}_\text{min}$ to be measured, shown as a function of $\omega_\text{m} t$. We set $g_0/\omega_\text{m}=1$, $\sigma/\omega_\text{m}=2^{-1/4}$, and $\alpha=0$. 
The initial value of the two eigenvalues are $g_0$ and zero. There is a jump in these values when $\omega_\text{m} t$ becomes larger than zero, i.e., when the interaction is turned on. For large interaction times, 
the eigenvalues tend to the same value $g_0$.}
\label{fig:eigenvM}
\end{figure}

The value of $\alpha$, the amplitude of the initial mechanical oscillator coherent state, has a strong influence on the value of $t^\ast{}$. We show in \fref{costfig} that the limit $\alpha\rightarrow\infty$, 
with $\alpha \in \mathbb{R}$, results in $t^\ast{}=0$ and the lowest observed value for $\bar C_{\text{min}}\approx0.636\sigma^2$. However, the eigenvalues of $\hat{M}^\ast_\text{min}$ are still zero and $g_0$ 
at $t=0$, from which it follows that highly excited initial states of the mechanical oscillator result in an estimation scenario where measuring $M^\ast_\text{min}$ merely reinforces prior knowledge and 
yields only the mean $g_0$ of the prior probability distribution $p(g)$. In the next step, we investigate the position of the minimum for $\alpha \in \mathbb{C}$, to deduce its dependence on the phase of $\alpha$. 
\Fref{costimalpha} shows a shift of $t^\ast{}$ towards higher values and an increase of the minimum value of $\bar C_{\text{min}}$ as the imaginary part of $\alpha$ gets larger. We see that the case with very 
large $\lvert\alpha\rvert$ may lead to inconclusive measurement scenarios because $\bar C_{\text{min}}$ is only significantly smaller than its maximum for a short time period. This observation is of significance 
in the discussion of initial thermal states, since it implies that higher initial temperatures will degrade the quality of the estimation procedure.

\begin{figure}
\includegraphics[width=0.45\textwidth]{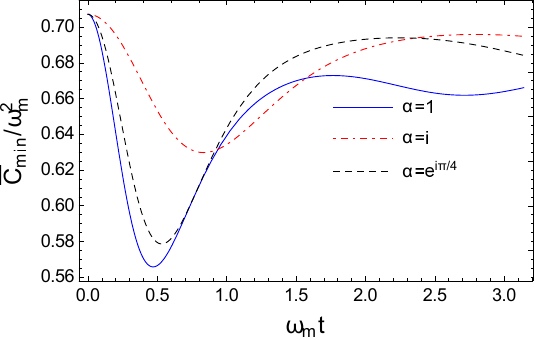}
\includegraphics[width=0.45\textwidth]{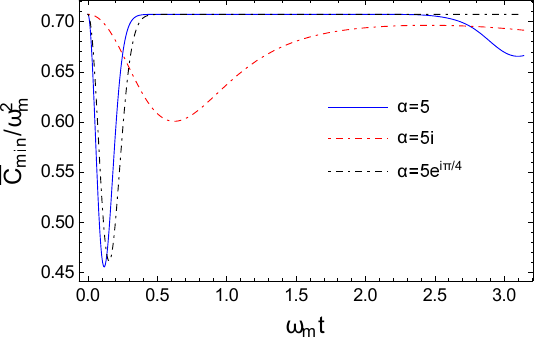}
\caption{The average minimum cost of error $\bar C_{\text{min}}/\omega^2_\text{m}$ as a function of $\omega_\text{m} t$. We set $g_0/\omega_\text{m}=1$ and $\sigma/\omega_\text{m}=2^{-1/4}$. The imaginary part 
of $\alpha$ shifts the value of the minimum to the right. Minima occurring at longer times are also less pronounced. Top:\ $\lvert\alpha\rvert=1$. Bottom:\ $\lvert\alpha\rvert=5$.}
\label{fig:costimalpha}
\end{figure}

Let us turn our attention to $\hat{M}^\ast_\text{min}$, which has already been defined as the optimal measurement, made at the time $t^\ast{}$ that minimizes the average minimum cost of error. Every outcome of 
the measurement of $\hat{M}^\ast_\text{min}$ is an estimate of $g$. The most important quantity for a possible experimental implementation is the average estimate at $t=t^\ast{}$
\begin{equation}
h(g)=\Tr\{\hat{M}^\ast_\text{min}\hat{\rho}_\text{F}(g)\}.
\end{equation}
Thus, measurement data determine the value of $h(g)$. From this, one may deduce the value of $g$. In \fref{avestimator}, we show the curves of $h(g)$ for different values of the real parameter 
$\alpha$. $t^\ast{}$ is independently calculated for each specific initial state. When $\alpha=0$, the average estimator is an even function of $g$. This is a direct consequence of our particular 
choice of the cost function~\eqref{eq:costf}, which is also an even function.

Before turning our attention to initial thermal states, let us conclude this section by summarizing the measurement procedure. Given a specific initial state, the system is allowed to evolve for a time $t^\ast{}$. 
At this point in time, one would conduct a measurement of $\hat{M}^\ast_\text{min}$. This process is repeated, obtaining an average measurement, $h(g)$. Using calculations of the kind shown in \fref{avestimator} 
allows one to work backward and obtain $g$.

\begin{figure}
\includegraphics[width=0.45\textwidth]{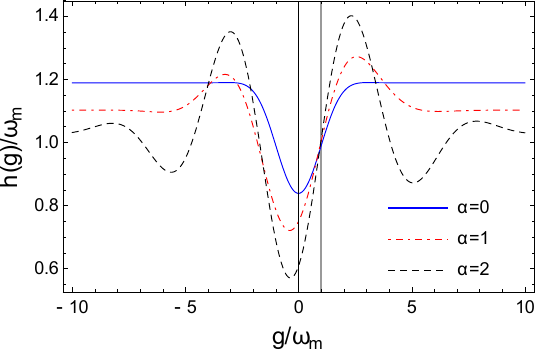}
\caption{The average estimator $h(g)/\omega_\text{m}$ as a function of $g/\omega_\text{m}$. We consider the amplitude $\alpha$ of the initial coherent state of the mechanical oscillator to be real. 
We set $g_0/\omega_\text{m}=1$ and $\sigma/\omega_\text{m}=2^{-1/4}$. The time is such that the average minimum cost of error $\bar C_{\text{min}}$ attains its minimum. The mean value $g_0$ of the prior 
probability distribution function $p(g)$ is depicted by a vertical line.}
\label{fig:avestimator}
\end{figure}

\subsection{Thermal state}

Similar to the case for an initial coherent state for the mechanical oscillator, an initial thermal state exhibits an average minimum cost of error $\bar C_{\text{min}}$ that is equal to $\sigma^2$ for $t=0$ 
and $t\to \infty$. In these limits, the eigenvalues of $\hat{M}_\text{min}$ have the same values as for the coherent state, so our earlier observations hold for the present case as well. In \fref{thermal}, we 
show the time dependence of $\bar C_{\text{min}}$ and the average estimator $h(g)$ for different average phonon numbers $n_{\text{th}}$ obtained from the density matrix~\eqref{eq:densitym} with the help of the 
expressions in \eref{example2}. One can observe that an increase in the value of $n_{\text{th}}$ increases $\bar C_{\text{min}}$ for most times, while it does not induce any significant change in the average 
estimator
$h(g)$. Furthermore, the oscillations seen in \fref{costfig} for longer interaction times are damped by the increase of $n_{\text{th}}$. Thus, in the context of this optimal estimation scenario the lower the 
temperature $T$ of the mechanical oscillator, the less sensitive is the average minimum cost of error. This provides additional impetus to one of the central pillars of optomechanical experiments, which is to 
cool down the mechanical oscillator to temperatures as low as possible~\cite{Aspelmeyer2014}.

\begin{figure}
\includegraphics[width=0.45\textwidth]{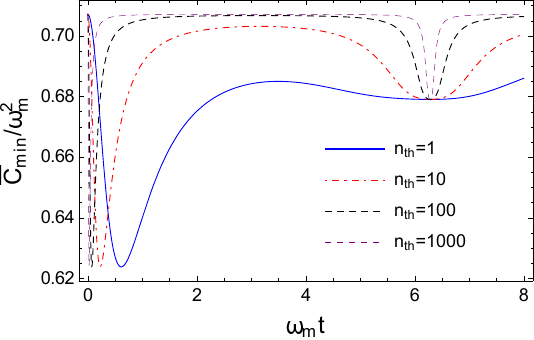}
\includegraphics[width=0.45\textwidth]{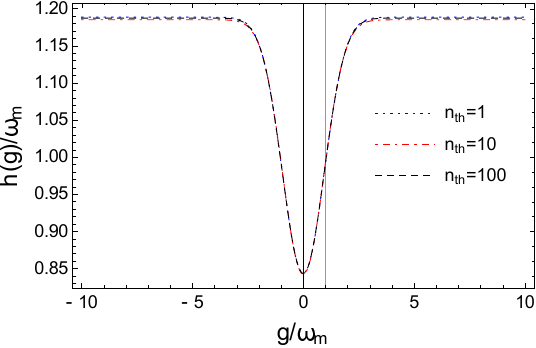}
\caption{Top: The average minimum cost of error $\bar C_{\text{min}}/\omega^2_\text{m}$ as a function of $\omega_\text{m} t$. We set $g_0/\omega_\text{m}=1$ and $\sigma/\omega_\text{m}=2^{-1/4}$. The oscillations 
at $\omega_\text{m} t\approx 2\pi$ are damped by the increase of the average phonon number $n_{\text{th}}$. Bottom: The average estimator $h(g)/\omega_\text{m}$ as a function of $g/\omega_\text{m}$. The time is 
such that the average minimum cost of error $\bar C_{\text{min}}$ attains its minimum. The mean value $g_0$ of the prior probability distribution function $p(g)$ is depicted by a vertical line.}
\label{fig:thermal}
\end{figure}

\subsection{Squeezed state}

We make use of \erefs{example3} to construct the density matrix in \eref{Aform}. The properties of $\bar C_{\text{min}}$ and $\hat{M}_\text{min}$ for $t=0$ and $t\to \infty$ are essentially the same as in the 
two cases discussed above. Let us recall that, in our discussion above, we showed that for an initial coherent state $\ket{\alpha}$ of the mechanical oscillator large $\lvert\alpha\rvert^2$ reduces the
average minimum cost of error $\bar C_{\text{min}}$, but at the expense of pushing the minimum towards very short interaction times, which leads to inconclusive measurement scenarios. In our discussion above we 
also identified a preferable scenario where $|\alpha|^2$ is large but with approximately equal real and imaginary parts. In the case of initial squeezed states, \fref{squeezed} shows an interesting effect, 
namely the squeezing parameter $\zeta$ may also reduce the average minimum cost of error. The average estimator $h(g)$ is again an even function, this time because we have chosen two squeezed states without 
displacement. The cost of error $\bar C_{\text{min}}$ attains a minimum when the squeezing angle lies between the position and momentum quadratures of the oscillator. This can be understood as an effective 
continuous sampling of the noise ellipse during the evolution of the system for the first fraction of a mechanical time-period. Squeezing along either position or momentum quadrature will result in a greater 
uncertainty, whereas squeezing at an angle half-way between these two quadratures allows the measurement to take place with the least possible uncertainty.

\begin{figure}
\includegraphics[width=0.45\textwidth]{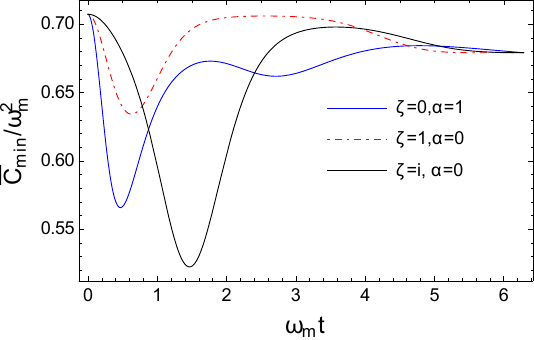}
\includegraphics[width=0.45\textwidth]{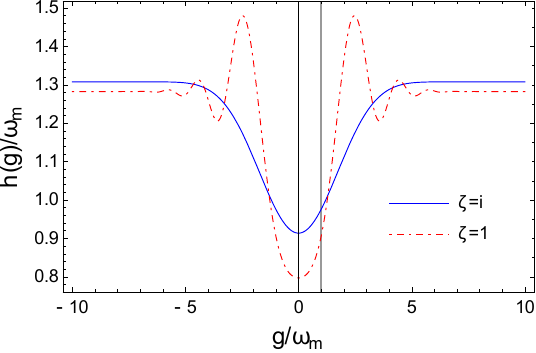}
\caption{Top: The average minimum cost of error $\bar C_{\text{min}}/\omega^2_\text{m}$ as a function of $\omega_\text{m} t$. We set $g_0/\omega_\text{m}=1$ and $\sigma/\omega_\text{m}=2^{-1/4}$. The case 
without squeezing, i.e, $\zeta=0$, has identical behavior to that in \fref{costfig}. Bottom: The average estimator $h(g)/\omega_\text{m}$ as a function of $g/\omega_\text{m}$ for $\alpha=0$. The time is such 
that the average minimum cost of error $\bar C_{\text{min}}$ attains its minimum. The mean value $g_0$ of the prior probability distribution function $p(g)$ is depicted by a vertical line.}
\label{fig:squeezed}
\end{figure}

\subsection{Different initial photonic states}

So far we have discussed in detail the estimation problem of the optomechanical coupling $g$ for the simplest initial state of the single-mode field. In this section, we consider the situation where the optical 
field may have more than one photon, and where the mechanical oscillator is initially in the ground state. Since $\hat{\rho}_\text{F}(g)$ depends on $g$ only in its off-diagonal elements, we therefore set the 
amplitude of all participating photon number states to be equal. This ensures the maximum allowed absolute value for the off-diagonal elements in the density matrix. Due to the added complexity of dealing with 
\eref{intx} we restrict our comparison to the following family of initial states of the optical field, indexed by the parameter $N=2,3,4$:
\begin{equation}
\ket{\psi_N}_\text{c}=\sum_{n=0}^{N-1}a_n \ket{n}_\text{c}= \frac{1}{\sqrt{N}}\sum_{n=0}^{N-1}\ket{n}_\text{c}.
\label{eq:initialstate}
\end{equation}
Our earlier investigations consider exclusively the case $N=2$.

\begin{figure}
\includegraphics[width=0.45\textwidth]{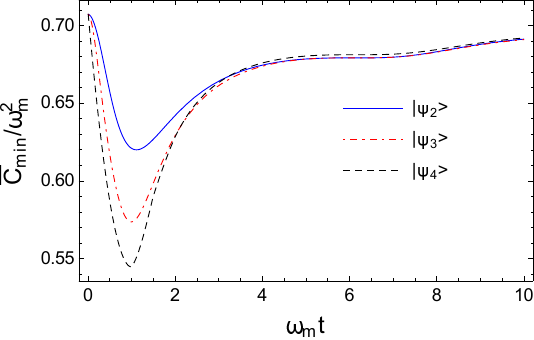}
\caption{The average minimum cost of error $\bar C_{\text{min}}/\omega^2_\text{m}$ as a function of $\omega_\text{m} t$ for different initial states of the optical field. We set $g_0/\omega_\text{m}=1$, 
$\sigma/\omega_\text{m}=2^{-1/4}$, and the mechanical oscillator initially in the ground state. $\ket{\psi_N}_\text{c}$, defined in \eref{initialstate}, is the initial optical state.}
\label{fig:costdim}
\end{figure}

\Fref{costdim} shows that the average minimum cost of error is reduced as the number of the photons in the initial state increases. This can be understood by examining carefully the Hamiltonian in \eref{optoH},
which reveals that the interaction between the single mode field and the mechanical oscillator gets stronger with increased number of participating photons. Thus, we have a better chance to estimate the 
optomechanical coupling $g$. The time $t^\ast{}$ when $\bar C_{\text{min}}$ attains its minimum does not change markedly with $N$. We have also calculated the average estimator $h(g)$ for $t^\ast{}$; 
\fref{estimatordim} shows the three curves obtained. Since the mechanical oscillator is initially in the ground state in every case, all curves are even.

\begin{figure}
\includegraphics[width=0.45\textwidth]{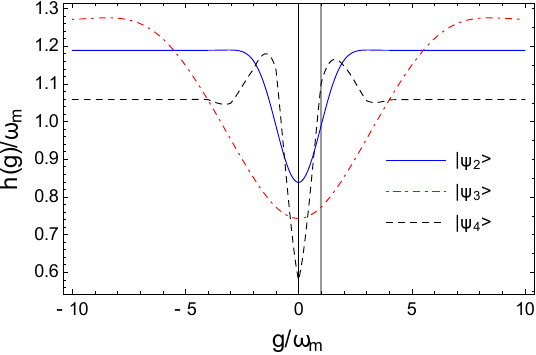}
\caption{The average estimator $h(g)/\omega_\text{m}$ as a function of $g/\omega_\text{m}$. We set $g_0/\omega_\text{m}=1$, $\sigma/\omega_\text{m}=2^{-1/4}$, and the mechanical oscillator initially in the 
ground state. The time is such that the average minimum cost of error $\bar C_{\text{min}}$ attains its minimum. The mean value $g_0$ of the prior probability distribution function $p(g)$ is depicted by a 
vertical line.}\label{fig:estimatordim}
\end{figure}

\section{Quantum Cram\'er--Rao-type inequality}
\label{sec:qCR}
In the preceding sections we discussed the properties of the optimum Hermitian operator $\hat{M}_{\text{min}}$ which minimizes the average cost in \eref{averagecost}, and the eigenvalues of which are the estimates 
of the unknown optomechanical coupling strength $g$. An important task is to find out the accuracy with which $g$ can be estimated. We would like to employ here the quantum Cram\'er--Rao inequality, which is 
widely used in the case of unbiased estimators~\cite{Helstrom1968, Helstrom1974}. In the present case, however, we have a biased estimator
\begin{equation}
 \Tr\bigl\{\hat{\rho}_\text{F}(g) (\hat{M}_{\text{min}}-g \hat{I})\bigr\}=f(g),
\label{one}
\end{equation}
where $f(g)$ is the bias of the estimation and is not necessarily equal to zero. To properly account for this situation, we have to review the derivation of the Cram\'er--Rao inequality.

Let us first, however, deal with an extra issue regarding the derivative of the density matrix $\hat{\rho}_\text{F}(g)$ with respect to the parameter $g$. For concreteness, let us recall the density matrix 
$\hat{\rho}_\text{F}(g)$ from \eref{densitym}, together with \erefs{example1}, and observe that
\begin{equation}
\hat{\rho}_\text{F}(g)=\sum^\infty_{n,m=0} a_n a^\ast_\text{m} e^{-a_1 (n-m)^2+a_2 (n^2-m^2)-a_3 (n-m)} \ket{n}\bra{m},
\end{equation}
where
\begin{align}
a_1=&\frac{g^2}{\omega^2_\text{m}} \bigl[1-\cos(\omega_\text{m} t) \bigr],\\
a_2=&i \frac{g^2}{\omega^2_\text{m}} \bigl[\omega_\text{m} t-\sin(\omega_\text{m} t) \bigr],\ \text{and}\\
a_3=&i \omega_\text{c} t-\frac{g}{\omega_\text{m}} \bigl[\alpha^\ast{}(1-e^{i\omega_\text{m} t})-\alpha(1-e^{-i\omega_\text{m} t}) \bigr].
\end{align}
Therefore,
\begin{align}
 \frac{\partial \hat{\rho}_\text{F}(g)}{\partial g}&=-\frac{\partial a_1}{\partial g}
 \bigl[\hat{a}^\dagger \hat{a},\bigl[\hat{a}^\dagger \hat{a},\hat{\rho}_\text{F}(g)\bigr]\bigr]\nonumber\\
&\qquad+\frac{\partial a_2}{\partial g} \bigl[(\hat{a}^\dagger \hat{a})^2,\hat{\rho}_\text{F}(g)\bigr]\nonumber\\
&\qquad-\frac{\partial a_3}{\partial g} \bigl[\hat{a}^\dagger \hat{a},\hat{\rho}_\text{F}(g)\bigr]=\mathcal{L}\bigl[\hat{\rho}_\text{F}(g)\bigr],
\label{eq:Liou}
\end{align}
which demonstrates that $\mathcal{L}\bigl[\hat{\rho}_\text{F}(g)\bigr]$ does not have the form of either a right logarithmic or a symmetrized logarithmic derivative of the density matrix 
$\hat{\rho}_\text{F}(g)$ by default (see the definitions in \aref{RSLD}). Therefore, we need the spectral decomposition of $\hat{\rho}_\text{F}(g)$ to construct at least the symmetrized logarithmic 
derivative operator, which is very challenging due to the fact that we have to deal with states defined on an infinite dimensional Hilbert space. Although this problem can be easily circumvented in 
numerical simulations, here we are motivated to derive an analytically expressible lower bound. This situation will result in a departure from the standard analysis~\cite{Helstrom1974}. In the standard proof, 
a Cauchy--Schwartz--Bunyakovsky inequality is employed, which suggests that in our new situation we would have to introduce the operator $\hat{\rho}^{-1/2}_\text{F}(g)$. This operator does not exist when the 
spectrum of $\hat{\rho}_\text{F}(g)$ contains zero (e.g., a pure state). We avoid this situation by following a different path.

In order to derive a lower bound for the mean-squared error,
\begin{equation}
 \mathrm{MSE}\bigl(\hat{M}_{\text{min}}\bigr)=\Tr\Bigl\{\hat{\rho}_\text{F}(g) (\hat{M}_{\text{min}}-g \hat{I})^2\Bigr\},
\end{equation}
we define
\begin{equation}
 x_1(g)=\Tr\bigl\{\hat{\rho}^2_\text{F}(g) \hat{M}_{\text{min}}\bigr\},
\end{equation}
and then we differentiate both sides with respect to the parameter $g$,
\begin{equation}
 \Tr\Biggl\{\Biggl(\frac{\partial \hat{\rho}_\text{F}(g)}{\partial g}\hat{\rho}_\text{F}(g)+\hat{\rho}_\text{F}(g)
 \frac{\partial \hat{\rho}_\text{F}(g)}{\partial g}\Biggr)
 \hat{M}_{\text{min}}\Biggr\}=x'_1(g).
\label{eq:two}
\end{equation}
We also define $x_2(g)=\Tr\bigl\{\hat{\rho}^2_\text{F}(g)\bigr\}$, and find
\begin{equation}
 \Tr\Biggl\{\Biggl(\frac{\partial \hat{\rho}_\text{F}(g)}{\partial g}\hat{\rho}_\text{F}(g)+\hat{\rho}_\text{F}(g)
 \frac{\partial \hat{\rho}_\text{F}(g)}{\partial g}\Biggr) g \hat{I}\Biggr\}=g x'_2(g).
\label{eq:three}
\end{equation}
Subtracting \eref{three} from \eref{two}, we obtain
\begin{multline}
 \Tr\Biggl\{\Biggl(\frac{\partial \hat{\rho}_\text{F}(g)}{\partial g}\hat{\rho}_\text{F}(g)+\hat{\rho}_\text{F}(g)
 \frac{\partial \hat{\rho}_\text{F}(g)}{\partial g}\Biggr) (\hat{M}_{\text{min}}-g \hat{I})\Biggr\}\\
=x'_1(g)-g x'_2(g)=x(g),
\label{eq:four}
\end{multline}
where the last equality defines the function $x(g)$. We make use of \eref{Liou} and write \eref{four} as
\begin{equation}
\Tr\bigl\{\bigl(\mathcal{L}\bigl[\hat{\rho}_\text{F}\bigr]\hat{\rho}_\text{F}+\hat{\rho}_\text{F}
 \mathcal{L}\bigl[\hat{\rho}_\text{F}\bigr]\bigr) (\hat{M}_{\text{min}}-g \hat{I})\bigr\}=x(g),
\label{eq:fh}
\end{equation}
where for the sake of notational simplicity we have omitted the argument $g$ of $\hat{\rho}_\text{F}(g)$.

\begin{figure}
\includegraphics[width=0.45\textwidth]{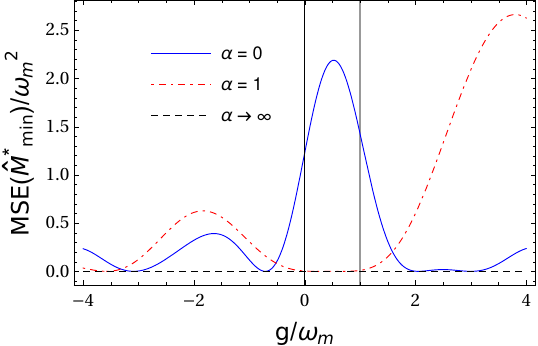}
\includegraphics[width=0.45\textwidth]{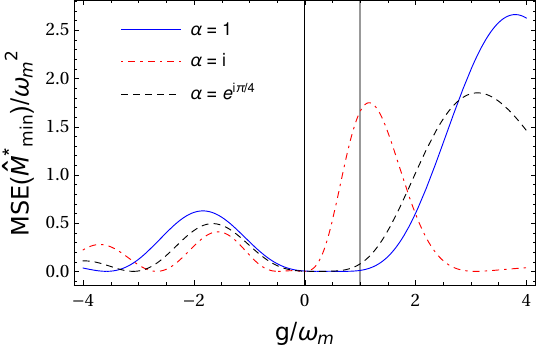}
\includegraphics[width=0.45\textwidth]{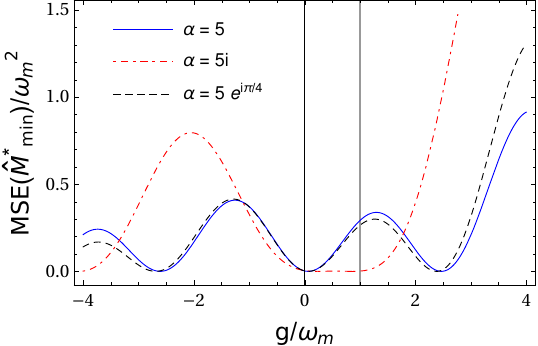}
\caption{The lower bound of the mean-squared error as a function of $g/\omega_\text{m}$. We set $g_0/\omega_\text{m}=1$ and $\sigma/\omega_\text{m}=2^{-1/4}$. The time is such that the 
average minimum cost of error $\bar C_{\text{min}}$ attains its minimum. Compare the top figure with \fref{costfig}, and the bottom two with \fref{costimalpha}. See also \fref{avestimator} for the corresponding 
average estimator.}
\label{fig:cr1}
\end{figure}

Before continuing, we discuss an issue connected with the boundedness of $\hat{\rho}_\text{F}$. The Banach space of the Hilbert--Schmidt operators is defined as
\begin{equation}
\mathcal{B}_2(\mathcal{H}):=\Bigl\{\hat X \in \mathcal{B}(\mathcal{H}):\, \Tr\{\hat X^\dagger\hat X\}<\infty\Bigr\},
\end{equation}
where $\mathcal{B}(\mathcal{H})$ is Banach space of all bounded operators defined on the Hilbert space $\mathcal{H}$. The space $\mathcal{B}_2(\mathcal{H})$ with the inner product
\begin{equation}
 \langle A, B \rangle=\Tr\{A^\dagger B\},
\end{equation}
where $A,B \in \mathcal{B}_2(\mathcal{H})$, is a Hilbert space~\cite{Reed1972}. The Cauchy--Schwartz--Bunyakovsky inequality reads
\begin{equation}
 \lvert\Tr\{A^\dagger B\}\rvert^2\leqslant \Tr\{A^\dagger A\}\Tr\{B^\dagger B\}.
\label{eq:inner}
\end{equation}
In our case the Hilbert space is the symmetric Fock space, i.e., $\mathcal{H}=\Gamma_\text{s}(\mathbb{C})$, and $\mathcal{L}$ contains powers of $\hat{a}^\dagger \hat{a}$, which is an unbounded operator. 
This clearly shows that our proof is limited to density matrices which fulfill the conditions $\hat{\rho}^{1/2}_\text{F} (\hat{a}^\dagger \hat{a})^2 \hat{\rho}_\text{F}, 
\hat{\rho}^{1/2}_\text{F} \hat{a}^\dagger \hat{a} \hat{\rho}_\text{F}\hat{a}^\dagger \hat{a} \in \mathcal{B}_2\bigl(\Gamma_\text{s}(\mathbb{C})\bigr)$. These conditions, together with the cyclic property 
of the trace, imply that $\hat{\rho}^{1/2}_\text{F} \mathcal{L}\left(\hat{\rho}_\text{F}\right)$ is a Hilbert--Schmidt operator. Similarly, the condition $\hat{\rho}^{1/2}_\text{F} \hat{M}_{\text{min}} 
\in \mathcal{B}_2\bigl(\Gamma_\text{s}(\mathbb{C})\bigr)$ may restrict further the set of the density matrices. In other words, there are restrictions on the choice of the $a_n$ in the 
initial state \eref{initial}. In the case of finite dimensional examples, i.e, if there exists an $N>0$ such that $a_n=0$ for $n \geqslant N$, these complications do not arise, because all matrices are 
Hilbert--Schmidt operators. This is the typical case encountered in numerical simulations.

\begin{figure}
\includegraphics[width=0.45\textwidth]{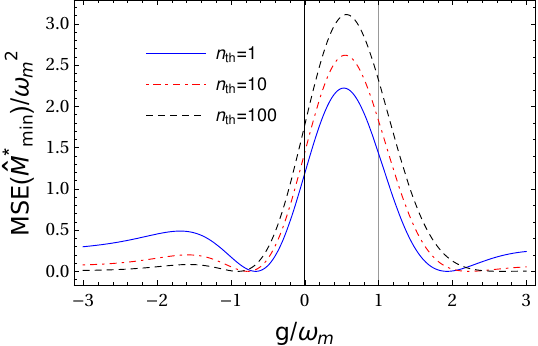}
\caption{The lower bound of the mean-squared error as a function of $g/\omega_\text{m}$. The curves were evaluated using the method and the parameters of \fref{thermal}.}
\label{fig:thq}
\end{figure}

\begin{figure}
\includegraphics[width=0.45\textwidth]{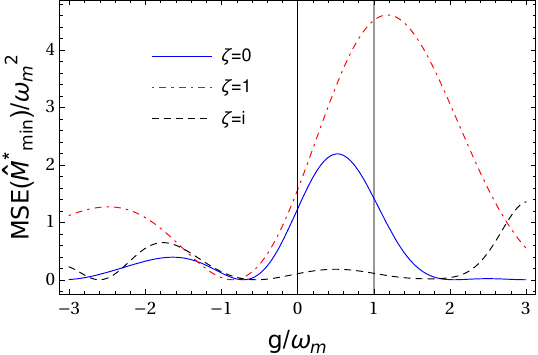}
\caption{The lower bound of the mean-squared error as a function of $g/\omega_\text{m}$. The curves were evaluated using the method and the parameters of \fref{squeezed}.}
\label{fig:sqq}
\end{figure}

\begin{figure}
\includegraphics[width=0.45\textwidth]{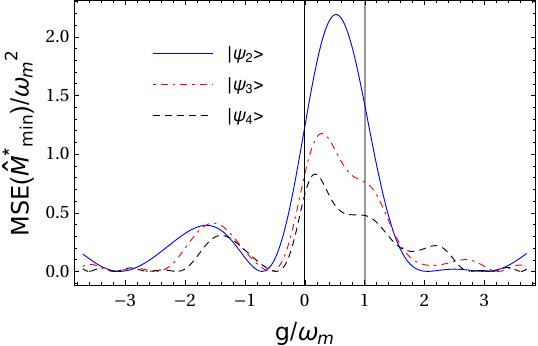}
\caption{The lower bound of the mean-squared error as a function of $g/\omega_\text{m}$. The curves were evaluated using the method and the parameters of \fref{estimatordim}.}
\label{fig:cr2}
\end{figure}

Now, provided that $\hat{\rho}^{1/2}_\text{F} \mathcal{L}\bigl[\hat{\rho}_\text{F}\bigr]$ and $\hat{\rho}^{1/2}_\text{F} \hat{M}_{\text{min}}$ are Hilbert--Schmidt operators, \eref{fh} implies
\begin{multline}
 \lvert{}x(g)\rvert=\Bigl|\Tr\Bigl\{\mathcal{L}(\hat{\rho}_\text{F})\hat{\rho}^{1/2}_\text{F} \hat{\rho}^{1/2}_\text{F}(\hat{M}_{\text{min}}-g \hat{I})\Bigr\}\\
 +\Tr\Bigl\{\hat{\rho}^{1/2}_\text{F} \mathcal{L}(\hat{\rho}_\text{F}) (\hat{M}_{\text{min}}-g \hat{I})\hat{\rho}^{1/2}_\text{F}\Bigr\} \Bigr|.
\end{multline}
Applying first the subadditivity of the absolute value and then the Cauchy--Schwartz--Bunyakovsky inequality~\eqref{eq:inner} twice, we find
\begin{equation}
 \lvert{}x(g)\rvert^2 \leqslant 2\Tr\Bigl\{\hat{\rho}_\text{F} \bigl(\mathcal{L}\bigl[\hat{\rho}_\text{F}\bigr]\bigr)^2 \Bigr\} \mathrm{MSE}\bigl(\hat{M}_{\text{min}}\bigr),
\end{equation}
where we have used the fact that $\bigl(\mathcal{L}\bigl[\hat{\rho}_\text{F}\bigr]\bigr)^\dagger=\mathcal{L}\bigl[\hat{\rho}_\text{F}\bigr]$, as can be deduced from \eref{Liou}.

Finally, we obtain a lower bound for the mean-squared error
\begin{equation}
\mathrm{MSE}\bigl(\hat{M}_{\text{min}}\bigr) \geqslant
\frac{\lvert{}x(g)\rvert^2}{2\Tr\Bigl\{\hat{\rho}_\text{F} \bigl(\mathcal{L}\bigl[\hat{\rho}_\text{F}\bigr]\bigr)^2 \Bigr\}}.
\label{eq:lbound}
\end{equation}
The quantity on the right of this inequality is very similar to the standard quantum Cram\'er--Rao bound. In this expression, the function of $x(g)$ in the numerator represents the fact that the estimator is 
biased, and includes information about the purity of the density matrix $\hat{\rho}_\text{F}(g)$. The denominator has a similar but slightly more complicated structure than the quantum Fischer 
information~\cite{Braunstein1994}, due to our approach to finding the derivative of $\hat{\rho}_\text{F}(g)$ with respect to the optomechanical coupling $g$.

We shall now apply the technique we just described to study the same cases as we did above, allowing us to study how the the mean-squared error behaves in each case.

\subsection{Coherent state}
We investigate numerically the lower bound of the mean-squared error. We consider $g_0$ and $\sigma$ to be the same as in \erefs{mst}. In \fref{cr1}, we recall the results of \frefs{costfig}
and~\ref{fig:costimalpha}, and show the behavior of the lower bound of the mean-squared error as a function of $g/\omega_\text{m}$. The most interesting feature occurs when $\lvert\alpha\rvert$ grows, where the lower 
bound is the smallest. This may seem to suggest that measurement strategies perform better under these conditions. However, this in apparent contrast with our findings in \sref{OM}. What we can 
deduce is that measurements made with large $\lvert\alpha\rvert$ may simply return $g_0$, i.e., our prior expectation, for the value of the coupling strength. In such circumstances, we gain no information 
about the system; these scenarios are therefore to be avoided.

\subsection{Thermal state}
Let us consider again the parameters of \fref{thermal}, where we have seen that the average estimator is insensitive to the change of the average phonon number $n_{\text{th}}$, i.e., the change in the 
temperature of the mechanical oscillator. Here, we observe something different for the lower bound of the mean-squared error, (see \fref{thq}). These findings indicate that the accuracy of the 
measurements is slightly worsened with the increase of $n_{\text{th}}$. However, we have also demonstrated an increase in the average minimum cost of error as $n_{\text{th}}$ is increasing. Therefore, 
in accordance with intuition, high temperatures once again lead to inconclusive estimation results.

\subsection{Squeezed state}
As we have seen in \fref{squeezed} for the average minimum cost of error, squeezing is beneficial in the sense of reducing the mean-squared error. In \fref{sqq} we see the lower bound of the mean-squared error 
may also be reduced by squeezing, in a manner that depends highly on the squeezing angle as well as the magnitude of the squeezing. This is, once again, in accordance with our earlier arguments and 
with intuition.

\subsection{Different initial photon states}
Finally, we compare the lower bound of the mean-squared error for the three different initial single-mode field states given in \eref{initialstate}, with the mechanical oscillator again assumed to be in 
its ground state. The time $t^\ast{}$, when the average minimum cost of error $\bar C_{\text{min}}$ attains its minimum is taken to be the same as in \fref{estimatordim}. The lower bound of the mean-squared error, 
as shown in \fref{cr2}, generally decreases with the photon number states in the initial state. This is in agreement with our findings in \fref{costdim}, namely that the average minimum cost of 
error is reduced by the increase of the photon number states. This also suggests that the initial preparation of the optical field is crucial to the outcome of the estimation procedure, with an equally 
weighted superposition of many photon number states being preferable. For ranges of values of $g$, however, either this improvement with increasing $N$ is not seen, or in some cases, the situation worsens 
as $N$ increases.

We conclude from this qualitative assessment that preparation of initial states of both the optical field and mechanical oscillator is crucial to obtaining more precise measurement outcomes, and 
consequently better estimations of the optomechanical coupling strength.

\section{Concluding remarks}
\label{sec:conclusion}
In this paper we have investigated the simplest optomechanical model, consisting of a single mode of an optical field interacting with a single vibrational mode of a mechanical oscillator, from the 
perspective of quantum estimation theory. The object of our analysis was to determine the optomechanical coupling strength optimally, based on measurements made on the optical field. We have discussed 
this problem by introducing a quantum estimation scenario, in which one seeks for the best estimator, which minimizes the mean-square error cost functional. This Bayesian-inference approach requires a 
prior probability density function of the coupling strength, which represents the limited prior information held about the system. In particular, we have considered a normal distribution, where the 
mean and the standard deviation have been set to values emerging from the derivation of the radiation pressure Hamiltonian~\cite{Law1995}. This derivation motivates our analysis, which develops an 
estimation procedure that results in a updated posterior probability density function for the coupling strength.

We have concentrated on the average mean-square error estimator, where the measurements occur at those interaction times where the average minimum cost of error reaches a minimum. The estimates are the 
eigenvalues of this estimator, with the eigenvectors determining a projective POVM that implements the measurement strategy. Our analysis has shown that highly excited initial coherent states of the 
mechanical oscillator limit the efficiency of this estimation procedure, unless the imaginary and the real parts of the displacement amplitude are approximately equal. We have demonstrated that the most 
promising estimates involve measurements being made during the first time period of the mechanical oscillation. We have, moreover, explored the effect of increasing the photon number states involved in the 
state of the optical field, sticking to the case of an equally weighted superposition of photon number states, \eref{initialstate}; we find that increasing photon numbers reduces the average information loss. 
Furthermore, we have investigated scenarios where the mechanical oscillator is initially in a thermal state or a squeezed state. In general, thermal states lead to inconclusive measurement outcomes, 
where the updated posterior probability density function is the same as the prior one. The situation with an initial squeezed state is different, because we find that for certain choices of squeezing angle, 
squeezing reduces the average minimum cost of error.

Third, we have investigated the accuracy of the mean-square error estimator by means of a lower bound for the mean-squared error. The quantum Cram\'er--Rao inequality, defining this lower 
bound, is derived with the help of a symmetrized logarithmic derivative operator. In our situation, this operator is demanding to construct due to the infinite dimensionality of the Hilbert space on which 
the states to be measured are defined. Therefore, we have derived a new lower bound, \eref{lbound}, for the mean-squared error of our biased estimator. In fact, we have reproduced the derivation of the 
quantum Cram\'er--Rao inequality by applying its standard methods to our case. Our numerical investigations here largely corroborate our previous conclusions. However, the lowest bounds for the estimation 
accuracy have been found for those limiting cases when the eigenvalues of the estimator are either zero or the mean of the prior normal distribution, where measurement yields no further information about the 
system. In particular, we have found that the initial state of the mechanical oscillator has to be carefully prepared, otherwise the outcome of the measurement process will be to simply reinforce prior 
expectations of the optomechanical coupling strength.

Finally let us make some comments on our approach. The analysis clearly indicates a characteristic set of parameters when the estimation of the optomechanical coupling can be done with minimal loss of 
information. Despite the fact that our results pinpoint some important results for a scenario of much experimental relevance, the question of how to \emph{implement} the optimal detection strategy or to 
compare with less optimal but implementable measurement setups (see Ref. \cite{Seveso2017}) has not been answered, and is the subject of ongoing investigations. Another critical point is the preparation 
of the initial state of the optical field; in this paper we have considered this state to be an equally-weighted superposition of photon number states, but we have not tackled the question of whether this 
family of states is optimal. These questions define the direction of our future investigations. As a final word we think that the present paper may offer interesting perspective and viewpoint, which 
provides a different way of thinking about optomechanical systems.

\section*{Acknowledgement}
J.Z.B.\ is grateful to Matteo G. A.\ Paris for stimulating discussions. This paper is supported by the European Union's Horizon 2020 research and innovation program under Grant Agreement 
No.\ 732894 (FET Proactive HOT--–Hybrid Optomechanical Technologies).

\appendix
\begin{widetext}
\section{Mechanical oscillator in an initial squeezed state}
\label{sec:Calc}
In this appendix, we present the full expressions of $I^{(0)}$, $I^{(1)}_{n,m}$, and $I^{(2)}_{n,m}$, which appear in \eref{example3}. First, we introduce the following notation:
\begin{align}
\chi^{(1)}_{n,m}&=n(e^{-i\omega_\text{m} t}-1)-m(e^{i\omega_\text{m} t}-1),\\
\chi^{(2)}_{n,m}&=n(e^{-i\omega_\text{m} t}-1)+m(e^{i\omega_\text{m} t}-1),\\
\chi^{(3)}_{n,m}&=n(1-e^{-i\omega_\text{m} t})e^{i(\theta-\omega_\text{m} t)}-m(1-e^{i\omega_\text{m} t})e^{-i(\theta-\omega_\text{m} t)},\ \text{and}\\
\chi^{(4)}_{n,m}&=n(1-e^{-i\omega_\text{m} t})e^{i(\theta-\omega_\text{m} t)}+m(1-e^{i\omega_\text{m} t})e^{-i(\theta-\omega_\text{m} t)},
\end{align}
as well as
\begin{align}
\xi^{(0)}&=\frac{1}{4\bigl[1-\tanh(\lvert\zeta\rvert)\cos(z)\bigr]},\\
\xi^{(1)}&=\frac{1-\tanh(\lvert\zeta\rvert)\cos(z)}{4\bigl[1-\tanh^2(\lvert\zeta\rvert)\bigr]},\ \text{and}\\
\xi^{(2)}&=i\frac{\tanh(\lvert\zeta\rvert)\sin(z)}{1-\tanh(\lvert\zeta\rvert)\cos(z)},
\end{align}
with $z=\theta-2\omega_\text{m} t$. Finally, we can write
\begin{align}
I^{(2)}_{n,m}&=\xi^{(0)}\bigl[\chi^{(1)}_{n,m}+\tanh^2(\lvert\zeta\rvert)\chi^{(3)}_{n,m}\bigr]^2-\xi^{(1)}\Bigl\{\chi^{(2)}_{n,m}+\tanh^2(\lvert\zeta\rvert)\chi^{(4)}_{n,m}+\xi^{(2)}\bigl[\chi^{(1)}_{n,m}+
\tanh^2(\lvert\zeta\rvert)\chi^{(3)}_{n,m}\bigr]\Bigr\}^2,\\
I^{(1)}_{n,m}&=4|\alpha|\Bigl(\xi^{(0)}\bigl[\chi^{(1)}_{n,m}+\tanh^2(\lvert\zeta\rvert)\chi^{(3)}_{n,m}\bigr]i\bigl[\sin(z_1)+\tanh(\lvert\zeta\rvert)\sin(z_2)\bigr]+\xi^{(1)}\bigl\{\chi^{(2)}_{n,m}+
\tanh^2(\lvert\zeta\rvert)\chi^{(4)}_{n,m}+\xi^{(2)}\bigl[\chi^{(1)}_{n,m}\nonumber\\
&\qquad+\tanh^2(\lvert\zeta\rvert)\chi^{(3)}_{n,m}\bigr]\bigr\}\bigl\{\cos(z_1)+\tanh(\lvert\zeta\rvert)\cos(z_2)-i\xi^{(2)}\bigl[\sin(z_1)+\tanh(\lvert\zeta\rvert)\sin(z_2)\bigr]\bigr\}\Bigr),\ \text{and}\\
I^{(0)}&=4|\alpha|^2\Bigl(\xi^{(0)}\bigl[\sin(z_1)+\tanh(\lvert\zeta\rvert)\sin(z_2)\bigr]^2+\xi^{(1)}\bigl\{\bigl[\cos(z_1)+\tanh(\lvert\zeta\rvert)\cos(z_2)\bigr]\nonumber\\
&\qquad-i\xi^{(2)}\bigl[\sin(z_1)+\tanh(\lvert\zeta\rvert)\sin(z_2)\bigr]\bigr\}^2\Bigr),
\end{align}
where $z_1=\omega_\text{m} t -\phi$ and $z_2=\omega_\text{m} t -\phi-\theta$.
\end{widetext}

\section{The symmetrized and the right logarithmic derivative operators}
\label{sec:RSLD}
In this appendix, we present some well-known material in order to support the arguments of this paper. In a single parameter estimation scenario, the symmetrized logarithmic derivative $\hat{L}$ of the
density matrix $\hat{\rho}(x)$ is defined by
\begin{equation}
\frac{\partial \hat{\rho}}{\partial x}=\tfrac{1}{2}(\hat{L}\hat{\rho}+\hat{\rho} \hat{L}).
\end{equation}
The operator $\hat{L}$ is Hermitian \cite{Helstrom1976}. If we consider the spectral decomposition 
\begin{equation}
\hat{\rho}=\sum_i p_i \ket{i} \bra{i},
\end{equation}
then
\begin{equation}
\hat{L}=\sum_{i,j}2\frac{\bra{i} \partial \hat{\rho}/\partial x \ket{j}}{p_i+p_j} \ket{i}\bra{j}
\end{equation}
satisfies the above definition. However, in order to construct $\hat{L}$ one must know the exact eigenvalues and the eigenvectors of $\hat{\rho}$.

Another way of defining the derivative of $\hat{\rho}(x)$ with respect to $x$ involves the non-Hermitian operator $\mathbb{L}$ that is the solution of the equation
\begin{equation}
\frac{\partial \hat{\rho}}{\partial x}= \hat{\rho}\mathbb{L}=\mathbb{L}^\dagger \hat{\rho},
\end{equation}
this being called the right logarithmic derivative operator. This operator may not exist for many density matrices, and in particular for those representing pure states~\cite{Helstrom1976}.

\end{document}